**The largest Forbush decrease in 20 years: Preliminary analysis of SEVAN network observations.**

*A. Chilingarian, T.Karapetyan, B.Sargsyan*

*Yerevan Physics Institute*

*Alikhanyan Brothers 2, Yerevan 36, Armenia*

Abstract

We present a preliminary analysis of the largest Forbush Decrease (FD) observed in solar cycle 25 using SEVAN network data. Triggered by consecutive Earth-directed CMEs on May 30 and 31, 2025, this event produced two distinct FD minima and a prolonged recovery, with significant anisotropies in cosmic ray response across the network. The timing of interplanetary shock arrivals was confirmed by SSC signatures at 05:22 UTC on June 1 and 10:19 UTC on June 2, marking the onset of each FD phase. The depth and temporal structure of the FD varied across SEVAN detectors, providing insights into the complex heliospheric and magnetospheric dynamics during this multi-CME event.

1. Introduction

The SEVAN network (Space Environment Viewing and Analysis Network), as part of the United Nations Basic Space Science (UNBSS) activities, was supported by the International Heliophysical Year 2007 (IHY 2007, Thompson et al., 2009) and the UN Office for Outer Space Affairs. Experts from the Cosmic Ray Division (CRD, Chilingarian et al., 2003) at the Yerevan Physics Institute developed a new hybrid particle detector that measures neutral and charged particles. The network's initial rollout included installations in Croatia, Bulgaria, and India (Chilingarian et al., 2009). Expansion continued with the installation of SEVAN detectors in Slovakia, Germany (Hamburg and Berlin), the Czech Republic, and atop Zugspitze, Germany's highest peak, in 2023 (Chilingarian et al., 2024a). Local SEVAN groups foster a research community in solar physics and high-energy atmospheric physics by analyzing neutrons and muons modulated during violent solar events and recording increased fluxes of electrons and gamma rays during thunderstorms and RREA/TGE events. Currently, 10 SEVAN detectors are operational in Armenia and the highest peaks of Eastern Europe and Germany.

Since the project's inception in 2007, the SEVAN network has provided valuable data on solar events throughout the highly active 23$^{rd}$ solar activity cycle (Chilingarian et al., 2018). During the relatively weak 24th solar activity cycle, the SEVAN network was extensively utilized for atmospheric physics research, uncovering new sources of natural radiation (Chilingarian et al., 2010; 2011; 2024a; Chum et al., 2020) and estimating maximum values of atmospheric electric fields (Chilingarian et al., 2021). The latest discovery was a significant increase in gamma radiation during winter snowstorms (Chilingarian et al., 2025).

With the maximum of the 25th solar activity cycle reached, SEVAN provides crucial information about the types of solar events and the energy of primary solar protons (Chilingarian et al., 2024c; 2024d). We have already published a brief report comparing magnetospheric and ground-level enhancement events from November 5, 2023, and May 11, 2024 (Chilingarian, 2024). Additionally, we review the first Forbush decrease (FD, Forbush, 1950) of the 25th solar cycle that occurred before 2024. Here, we present preliminary results from observations regarding the largest FD in the 25th cycle, which took place on June 1-2, just at the beginning of the seventeenth Workshop "Solar Influences on the Magnetosphere, Ionosphere, and Atmosphere" in Primorsko, Bulgaria.

2. Solar and Interplanetary sources

On May 29, 2025, a G3-level geomagnetic storm began, driven by a coronal hole high-speed stream (CH HSS) that rotated into a geoeffective position, sending fast solar wind toward Earth. As the stream moved forward, it interacted with a co-rotating interaction region (CIR), a compression zone between slow and fast solar wind flows, resulting in an unexpectedly strong geomagnetic disturbance.

On May 30, Active Region 4100 produced a significant M3.4-class solar flare at 12:53 UTC, launching a moderately fast halo CME directed toward Earth. This CME was preceded by an earlier CME launched at 06:38 UTC, and the two events likely merged during propagation, forming a compound ICME. The shock from this compound CME system was detected at 05:22 UTC on June 1 by the ACE and DSCOVR spacecraft at the L1 point. Its arrival was marked by a sharp increase in interplanetary magnetic field strength ($B_{total}$) from approximately 7 nT to above 24 nT, along with a surge in solar wind speed from approximately 700 km/s (already elevated due to the CH HSS) to over 1100 km/s. According to USGS, this SSC marked the onset of a severe geomagnetic storm, with Dst plunging to –122 nT by 10:07 UTC. The passage of this interplanetary shock initiated the first phase of the Forbush Decrease (FD), characterized by a sudden drop in cosmic-ray flux with a rather complicated structure.

On May 31, the Sun displayed increased activity, with five active regions on its Earth-facing hemisphere. AR4100 alone produced three M-class flares, the most powerful being an M8.2-class flare at 00:05 UTC. This eruption launched a rapid halo CME with an estimated speed of 1,900 km/s. The interplanetary shock from this CME reached Earth at 10:19 UTC on June 2, marked by another sharp increase in $B_{total}$ from 6.5 to 24 nT at speeds between 650 and 850 km/s. The arrival of this CME intensified the ongoing geomagnetic storm to G4-level severity. The fast May 31 CME likely overtook the earlier, slower CMEs from May 30, creating a complex interplanetary CME (ICME) structure. This event triggered the second, deeper phase of the Forbush Decrease, aligning with the second minimum in cosmic-ray flux.

The combination of multiple CMEs, some of which overtook each other on their way to Earth, resulted in the deepest Forbush Decrease in 20 years, marked by a highly structured and prolonged evolution. The recovery phase lasted nearly a week, and global particle flux measurements showed complex spatial-temporal patterns, indicating large-scale heliospheric structuring and persistent disturbances in the interplanetary magnetic field.

## 3. SEVAN network response to geomagnetic disturbances

Significant disturbances in the magnetosphere caused large variations in cosmic ray flux, as measured on Earth's surface by neutron monitors and SEVAN detectors. The two-phase Forbush decrease began on June 1 and was extended by successive ICMEs, prolonging the recovery period for a week. Figure 1 displays the 10-minute averaged count rates from SEVAN detectors with upper 5 cm thick scintillators, located on mountain peaks: Aragats, Armenia (40.25N, 44.15E, 2000 and 3200 m), Lomnicky Stit, Slovakia (49.2N, 20.22E, 2634 m), and Musala, Bulgaria (42.1N, 23.35E, 2930 m). Data from Mileshovka Hill (50.6N, 13.9E, 837 m) in the Czech Republic is also included. The first FD minimum related to the "cannibal" ICME detected at 05:22 UTC on June 1 shows a minimum of 7.5% consistently at all locations at approximately 15:00 on June 1. The second minimum associated with the M8.2-class flare triggered a rapid halo CME that reached Earth at 10:19 UTC on June 2. The SEVAN network indicates an anisotropy between the European and Aragats detectors. While the Aragats SEVAN shows shallower minimum, the European SEVAN deepens the particle flux to 7.9% on Lomnicky Stit and 8.8% at Musala at approximately 17:40 on June 2.

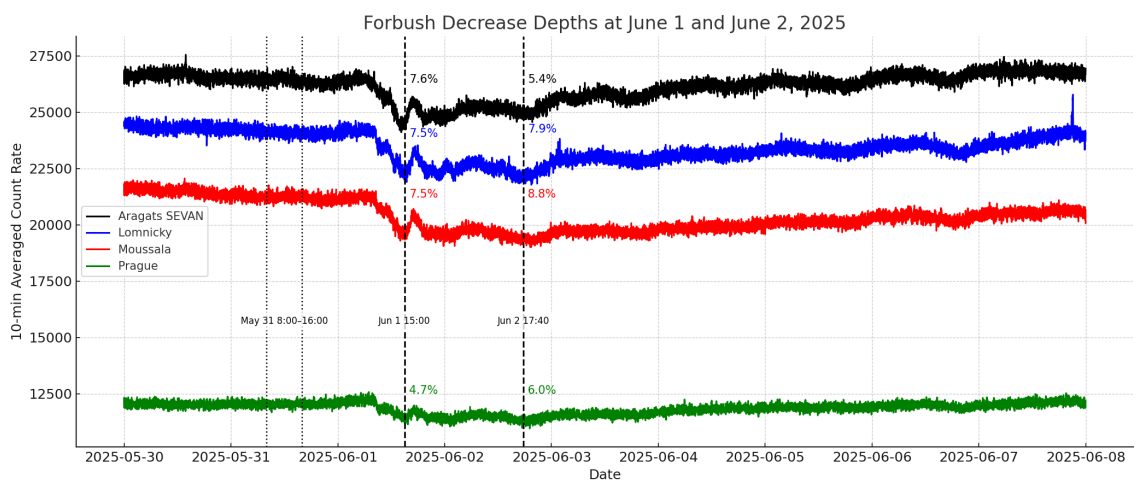

Figure 1. Forbush decrease observed by the upper scintillators SEVAN network. The depletion percentage was calculated relative to the mean values from 8:00 to

16:00 on May 31 (shown by dotted lines). Dashed lines show the deepest depletions of FD.

Figure 2 shows the same FD registered by the SEVAN detector, with coincidences selected from samples enriched by neutrons and muons. The Aragats neutron flux demonstrates a faster recovery than the European SEVANs. The Mileshovka neutrons show the deepest FD at 12%. In contrast, the muon flux (E > 200 MeV) from the Mileshovka SEVAN shows the smallest depletion compared to the mountain SEVANs.

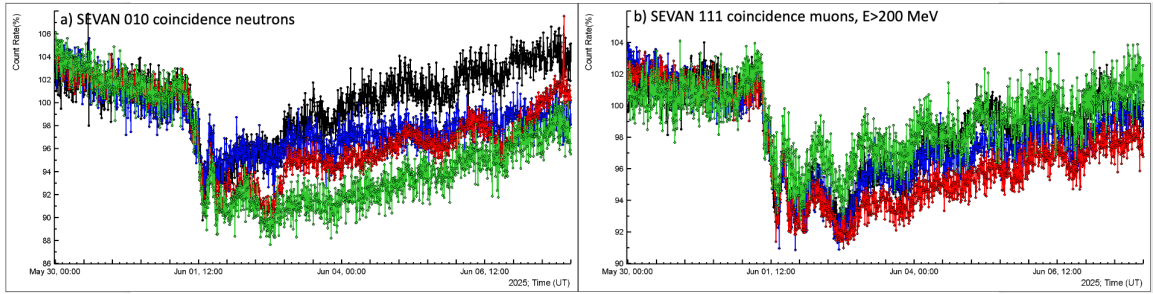

**Figure 2. Forbush decrease due to SEVAN coincidences involving selected neutrons and muons.** The depletion percentage was calculated relative to the mean values obtained from 8:00 AM to 4:00 PM on May 31.

Neutrons are more sensitive to FD particles. Figure 3 shows FD measured by Aragats (ArNM) and Nor Amberd (NANM) neutron monitors. Only the Aragats and Lomnicky Stit stations have neutron monitors and SEVAN detectors. The simultaneous detection of solar events by two types of detectors provides a basis for intercalibrating both networks. The ArNM shows a 15% depletion, while the NANM indicates a 2% smaller depletion due to atmospheric cutoff.

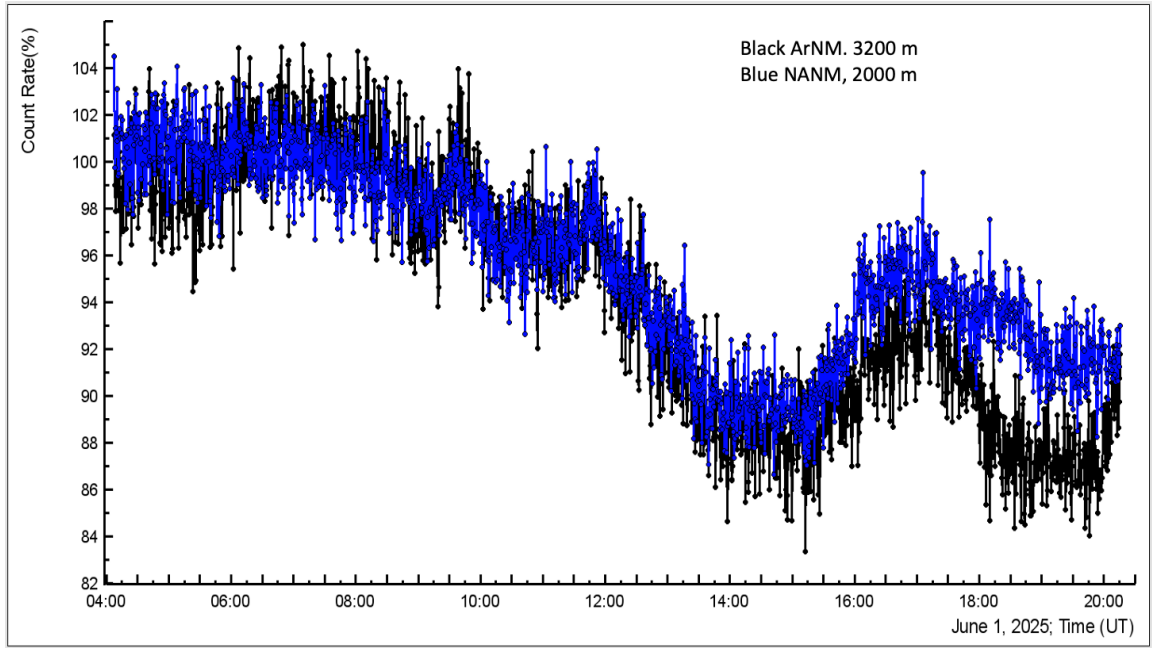

**Figure 3. Forbush decrease observed by ArNM (3200 m) and NANM (2000 m).**

Muon detectors with different energy thresholds are very useful for classifying solar events and assessing the primary solar proton energy at ground-level enhancements (GLEs) and magnetospheric effects (MEs) (Chilingarian et al., 2024b; 2024c). Figure 4 shows how the STAND3 detector's coincidences (Chilingarian and Hovsepyan, 2023) pictured the FD, which separates muons with different threshold energies. Most influenced by geomagnetic disturbances are 40 MeV muons, showing 12% depletion.

Another muon detector at Aragats, with energy thresholds of 300 and 400 MeV, shows only an 8% depletion. In Figure 4, we also display the thunderstorm ground enhancements (TGEs, Chilingarian et al., 2010, 2011) that occurred during excessive thunderstorms on May 30 and June 6. During thunderstorms, the atmospheric electric field accelerates free electrons. Under specific atmospheric conditions, these electrons form electromagnetic avalanches (cascades of secondary particles initiated by runaway electrons) that reach the surface and are registered as short particle bursts. Detectors with low energy thresholds effectively register such events.

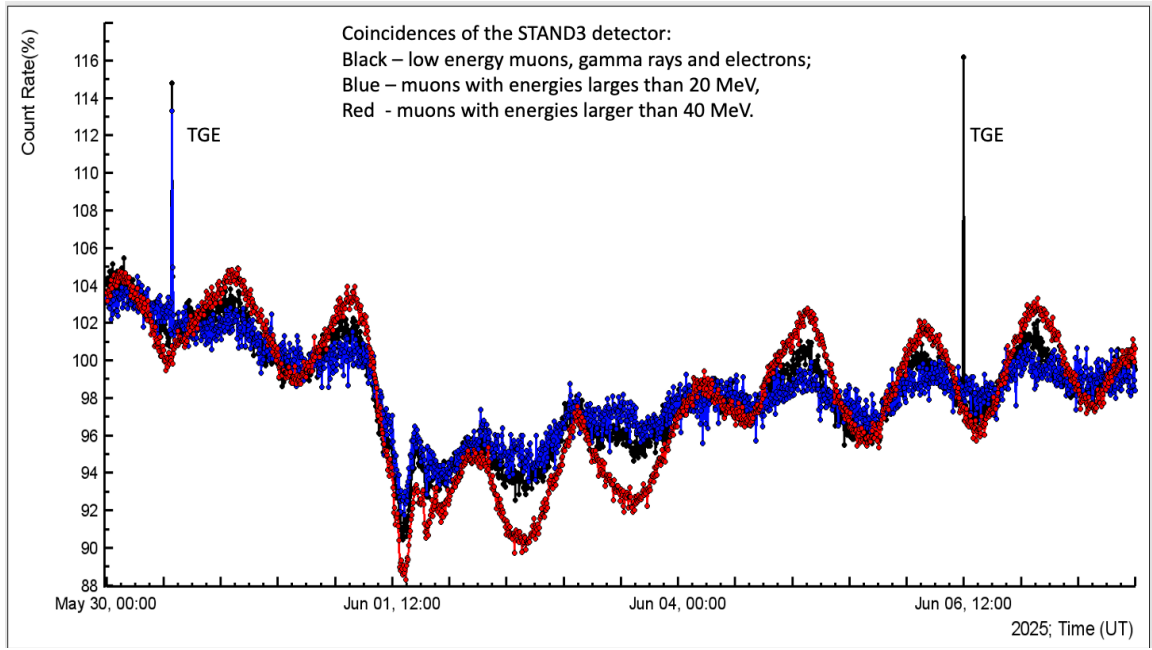

**Figure 4. FD registered by the stacked STAND3 detector. The vertical outburst in count rate shows TGEs – an impulse enhancement of count rates due to the operation of atmospheric electron accelerators.**

4. **Discussion and conclusions**

The combined neutron and muon observations by the SEVAN network provide complementary insights into the nature of the FD, capturing both global modulation and local dynamic effects.

The neutron count rate exhibited a smooth and gradual decrease, characteristic of global cosmic ray suppression by interplanetary magnetic structures. Neutrons are produced by spallation reactions when primary cosmic rays interact with atmospheric nuclei. Neutrons are extremely sensitive to Forbush Decreases because they originate from primary cosmic rays with energies around 1–20 GeV, where the modulation is strongest. Unlike charged particles, neutrons are not affected by magnetic fields once produced, making them an excellent probe of the primary cosmic ray flux variations.

Neutron monitors have historically been the primary instruments for detecting Forbush Decreases because they show large amplitude reductions (10–20% for strong events). In contrast, muon detectors, sensitive to higher-energy primaries (10–1000 GeV), reveal complementary information about both global and local geomagnetic modulation. Meanwhile, detectors sensitive to low-energy electrons and gammas display weak or unclear FD signals due to strong atmospheric and environmental background contamination.

The muon count rate showed a similar overall FD amplitude, with additional short-term spikes superimposed on the general trend. These muon spikes are interpreted as signatures of magnetospheric effects, likely caused by transient variations in the geomagnetic cutoff rigidity during the ongoing reconfiguration of the geomagnetic field.

Muons, being sensitive to higher-energy cosmic rays, respond to both large-scale heliospheric modulation and local geomagnetic changes, while neutrons primarily trace the broader suppression of cosmic rays. Forbush decreases have a greater impact on low-energy primary cosmic rays; however, the secondary electrons and gammas generated by these cosmic rays are less modulated, which dilutes the Forbush decrease signature in low-energy threshold detectors. The higher-threshold muon detectors, shielded from soft secondaries, observe a cleaner muon flux with a more pronounced relative Forbush decrease signature. Therefore, the 40 MeV threshold scintillator exhibits a larger amplitude of the Forbush decrease. These results underscore the importance of energy-differentiated muon channels in distinguishing Forbush decreases from overlapping magnetospheric effects and atmospheric secondaries.